\documentclass[%
 reprint,
 amsmath,amssymb,
 aps,
floatfix,
]{revtex4-2}
\usepackage{graphicx}% Include figure files
\usepackage{dcolumn}% Align table columns on decimal point
\usepackage{bm}
\usepackage{amsmath,amssymb,amsfonts}
\usepackage{algorithmic}
\usepackage{graphicx}
\usepackage{textcomp}
\usepackage{xcolor}
\usepackage{braket}
\usepackage{mathtools} 
\usepackage{qcircuit}% bold math
\begin{document}

\preprint{APS/123-QED}

\title{Investigating the Exchange of Ising Chains on a Digital Quantum Computer}

\author{Bassel Heiba Elfeky}
 \affiliation{Department of Physics, New York University, NY 10003, USA}
\author{Matthieu C. Dartiailh}
 \affiliation{Department of Physics, New York University, NY 10003, USA}
\author{S. M. Farzaneh}
 \affiliation{Department of Physics, New York University, NY 10003, USA}
\author{Javad Shabani}
 \affiliation{Department of Physics, New York University, NY 10003, USA}

\date{\today}

\begin{abstract}
The ferromagnetic state of an Ising chain can represent a two-fold degenerate subspace or equivalently a logical qubit which is protected from excitations by an energy gap. 
We study a a braiding-like exchange operation through the movement of the state in the qubit subspace which resembles that of the localized edge modes in a Kitaev chain. 
The system consists of two Ising chains in a 1D geometry where the operation is simulated through the adiabatic time evolution of the ground state.  
The time evolution is implemented via the Suzuki-Trotter expansion on basic single- and two-qubit quantum gates using IBM's Aer QASM simulator. 
The fidelity of the system is investigated as a function of the evolution and system parameters to obtain optimum efficiency and accuracy for different system sizes. Various aspects of the implementation including the circuit depth, Trotterization error, and quantum gate errors pertaining to the Noisy Intermediate-Scale Quantum (NISQ) hardware are discussed as well. 
We show that the quantum gate errors, i.e. bit-flip, phase errors, are the dominating factor in determining the fidelity of the system as the Trotter error and the adiabatic condition are less restrictive even for modest values of Trotter time steps. 
We reach an optimum fidelity $>99\%$ on systems of up to $11$ sites per Ising chain and find that the most efficient implementation of a single braiding-like operation for a fidelity above $90\%$ requires a circuit depth of the order of $\sim 10^{3}$ restricting the individual gate errors to be less than $\sim 10^{-6}$ which is prohibited in current NISQ hardware. 

\end{abstract}

\maketitle

\section{Introduction}

The race for developing fault-tolerant quantum computing has only intensified in the last few years \cite{rosenblum_fault-tolerant_2018,lescanne_exponential_2020}. Recent experiments show quantum hardware are beginning to show their advantage, although in very strict cases, over classical computers \cite{arute_quantum_2019}. While the main issue remains to be decoherence and local perturbations \cite{corcoles_challenges_2019,ozhigov_physicallimit_2019}, there has been significant progress in two directions. First, regarding qubits that are engineered to autonomously correct errors with methods such as bosonic codes \cite{chuang_bosonic_1997}, and second, with proposed qubits based on non-abelian anyons that exhibit non-trivial topological statistics,  which are immune to decoherence and local perturbations, such as Majorana qubits \cite{kitaev2003fault}. While these novel qubits are being developed, it is instructive can study their dynamics by simulating them on available noisy intermediate-scale quantum (NISQ) hardware.

Kitaev recognized that a 1D chain of spin-less fermions with superconducting pairing, known as a Kitaev chain, can host non-abelian anyons \cite{nayak_nonabelian_2008, kitaev_unpaired_2001}. Recently, signatures of non-abelian anyons have been claimed to be experimentally observed \cite{suominen_zero-energy_2017, mourik_signatures_2012}. The braiding, or series of exchanges, of Majorana fermions is the basis of topological quantum computing \cite{alicea_non-abelian_2011,litinski_quantum_2018, aasen_milestones_2016}.
While in Kitaev chains, the ground states' degeneracy is not lifted under the influence of local perturbations, these chains are difficult to realize. A more accessible model is the Ising chain. The mapping between the Kitaev chain and the Ising chain can be done using the Jordan-Wigner transformation. However, under such mapping topological protection is lost. Nonetheless, one might be able to study Majorana-like systems implemented on Ising chain based models \cite{backens_emulating_2017} as the mapping between the Ising Hamiltonian and the Kitaev Hamiltonian is exact and therefore the excitation gap is the same. While topological protection is lost, it has also been shown that some non-topological immunity to local noise and error can be achieved depending on the physical implementation, e.g., in photonic systems \cite{xu2016simulating}, and trapped ions \cite{mezzacapo2013topological}.
Moreover, due to the ferromagnetic order of the ground state of the Ising chain, one can see that even in the presence of excitations such as bit flips, the ground state can be recovered through a majority rule \cite{freeman2017engineering}.

 Simulations of braiding Majorana fermions in topological superconductors are discussed in \cite{narozniak_majorana_2020}. Further, dynamics of Majorana fermion braiding on a T-junction geometry using scaled two-qubit quantum gates on IBM quantum computers are investigated in \cite{stenger_simulating_2020}. Numerical simulations, as done in \cite{classical_simulation}, even when performed with braiding approximations such as weaving, proved to be inefficient. The simulation of Majorana fermions has also been studied as part of systems such as a cavity QED lattices in \cite{cavity_sim} but without performing braiding or operations on the modes. Recently, Backens et al.\cite{backens_emulating_2017} has shown how Majorana fermion braiding can be studied on Ising chains through fermionic mapping with a 1D geometry which is debatable given the non-topological nature of the system or particles. Here we study the system in terms of quantum gates. To simulate the quantum system in an appropriate environment practically and efficiently, we simulate the system on a quantum simulator running on a classical computer with the 1D geometry described in Backens et al. \cite{backens_emulating_2017} using basic quantum logic gates to investigate the implementation of such a braiding protocol. 

We provide a detailed analysis of the implementation of a braiding-like exchange protocol as well as the optimization and estimation of the resources required to achieve high fidelity on NISQ hardware. Even though \cite{backens_emulating_2017} and \cite{stenger_simulating_2020} claim to preform "braiding of Majorana zero modes" using similar Ising chain-based models, we use the term "braiding" loosely in this work since true braiding involves an operation in 2D physical space and involves an intrinsic notion of chirality. Chirality plays a major role in the braiding of Majorana zero modes and depends on local characteristics of the system and the exact procedure used to move the modes \cite{clarke_majorana_2011, sau_controlling_2011}. In our model, braiding could be thought as the usual 2D braiding mapped out onto a local 1D system, an exchange of two particles.

For the digital quantum simulations, we map our physical system onto a set of qubits and implement our simulation using IBM's open-source software package, Qiskit, and run the simulation on IBM's Aer QASM simulator. However, our approach can be applied to any digital quantum computer. Current NISQ era quantum hardware have limited circuit depth which can be seen as the number of non-simultaneous gates between the input and the output. We also attempt to minimize the circuit depth and estimate the minimal required resources for a given system.

For Hamiltonians that involve multiple non-commuting parts, there is no simple circuit representation of the system's evolution operator. To circumvent this issue, we will approximate the evolution using the Suzuki-Trotter decomposition, known as the Trotter expansion, described by Hatano and Suzuki \cite{Hatano_trotter_2015}. This expansion approximates the total evolution operator using the product of the evolution operator for each component over a short time $\Delta t$. Since this description is only an approximation, we have to choose the order of the expansion and the steps' duration to optimize the fidelity and reduce the circuit depth.

In this paper, we start by introducing the model and the braiding protocol in section II followed by the  digital Hamiltonian simulation in Section III. Then, in section IV, we discuss the system's expected Suzuki-Trotter error before presenting the results and analysis of the digital simulation and the contributing errors. 

\section{Ising Chain Implementation of Braiding-like Exchange Protocol}

\label{ising.states.ltr}
Majorana fermions have been predicted to exist in the 1D Kitaev model \cite{kitaev_unpaired_2001}, which consists of spinless fermions with an attractive interaction. This system can be mapped to an Ising chain through the Jordan-Wigner transformation \cite{greiter_1d_2014}. This mapping is exact, and the topological implications have equivalent counterparts in the spin space. However, the topological protection is lost in the transformation and the Ising chain is sensitive to local perturbations.

The Ising model describes a lattice of spins exhibiting nearest-neighbor interactions and interactions with an external transverse magnetic field. In the case of a homogeneous nearest-neighbor interaction, the Hamiltonian of the system can be written as \cite{pfeuty_ising_1970}:

\begin{equation}
    H_{Ising}=-J\sum\limits_{n=1}^{N-1}\sigma^{z}_{n}\sigma^{z}_{n+1}-\sum\limits_{n=1}^{N}h_{n}\sigma^{x}_{n}
    \label{eq:ising}
\end{equation}
    
where $J$ represents the strength of the interaction between neighboring spins and $h_{n}$ represents the strength of the magnetic field at site $n$. 

\begin{figure}[ht!]
    \centerline{\includegraphics[width=0.48\textwidth]{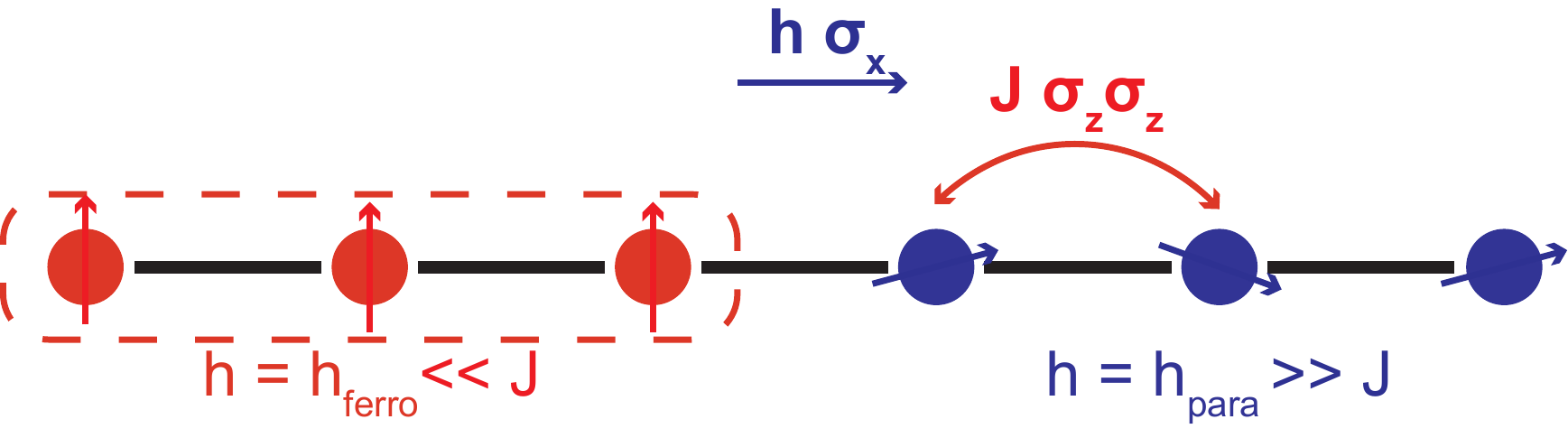}}
    \caption{An Ising chain with homogeneous nearest-neighbor interaction (J) and local transverse magnetic field (h). By controlling the transverse field's strength, the chain is split into two domains: left side is ferromagnetic ($h \ll J$), and right side is paramagnetic ($h \gg J$).}
    \label{fig_ising}
\end{figure}

Given a weak magnetic field compared to the strength of the pair interaction, $h_{n}<J$, the ground state of the system is ferromagnetic; when the opposite is true, $h_{n}>J$, the system is paramagnetic. To create well-defined ferromagnetic and paramagnetic domains on an Ising chain, we set the magnetic field, $h_{n}$, for each site $n$ as $h_{ferro}\ll J$ and $h_{para}\gg J$, respectively, as shown in Fig.~\ref{fig_ising}.

 In the case of a three-site chain, the two states, $\ket{\uparrow\uparrow\uparrow}$ and $\ket{\downarrow\downarrow\downarrow}$, of the ferromagnetic domain of an Ising chain, are degenerate and could span a qubit space. Consequently, the Ising chain's ferromagnetic domain can be used to mathematically represent the topological domain that hosts Majorana fermions \cite{backens_emulating_2017}. For the ferromagnetic states of the Ising chain the mapped states, which can be either be empty ($\ket{0}$) or occupied ($\ket{1}$), would then follow:

\begin{equation}
    \begin{aligned}
        \ket{0}=\frac{1}{\sqrt{2}}(\ket{\uparrow\uparrow\uparrow}+\ket{\downarrow\downarrow\downarrow})\\[3pt]
         \ket{1}=\frac{1}{\sqrt{2}}(\ket{\uparrow\uparrow\uparrow}-\ket{\downarrow\downarrow\downarrow})\\[3pt]
    \end{aligned}
    \label{eq:states}
\end{equation}
    
where $\ket{\uparrow\uparrow\uparrow}$ represents a ferromagnetic domain of length 3 with all its spins in an $\ket{\uparrow}$ state. Because of this mapping, we will now refer describe the topological domain as ferromagnetic domain.

\begin{figure}[ht!]
    \centerline{\includegraphics[width=0.48\textwidth]{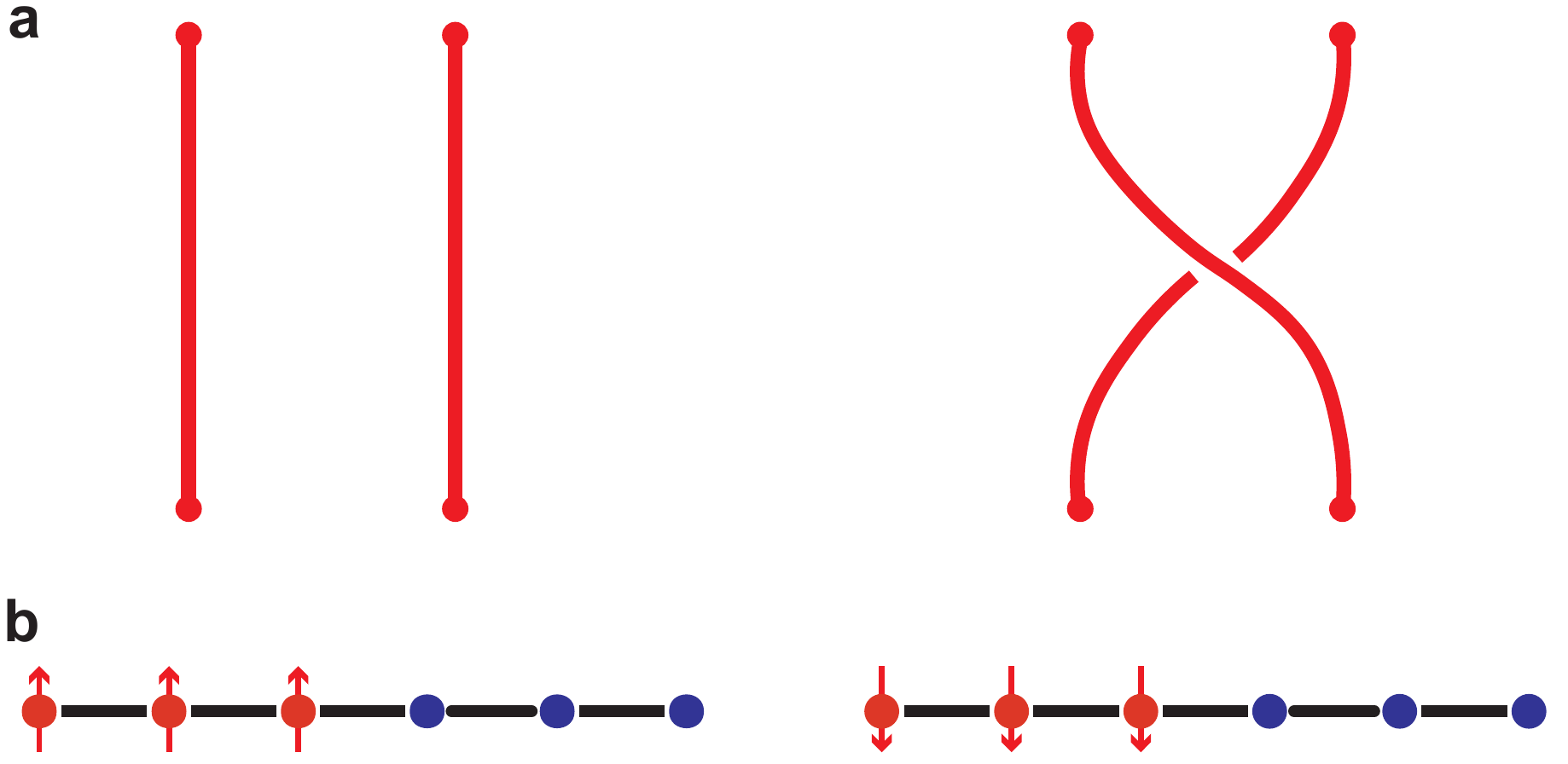}}
    \caption{Braiding of a Majorana-like pair with the corresponding Ising state representation. \textbf{a} Braiding of two Majorana quasiparticles in 2D \textbf{b} The evolution of the braiding can be represented using spins where the initial spin states of the left side is all up and the final state is all down.}
    \label{fig_spin-mapped-braiding}
\end{figure}
\label{ss:ising.braiding}
    
Braiding Majorana fermions requires preforming exchange in real space as seen in Fig.~\ref{fig_spin-mapped-braiding}a. Such an operation is strictly forbidden in 1D. When the Jordan-Wigner transformation is applied to multiple chains, issues arise that include non-local interactions and operator commutativity of different chains. To fix these issues, one can introduce an extra spin or coupler, $S^{\alpha}$, which ensures that the fermions of different chains properly anti-commute \cite{backens_emulating_2017}.

\begin{figure}[ht!]
    \centerline{\includegraphics[width=0.48\textwidth]{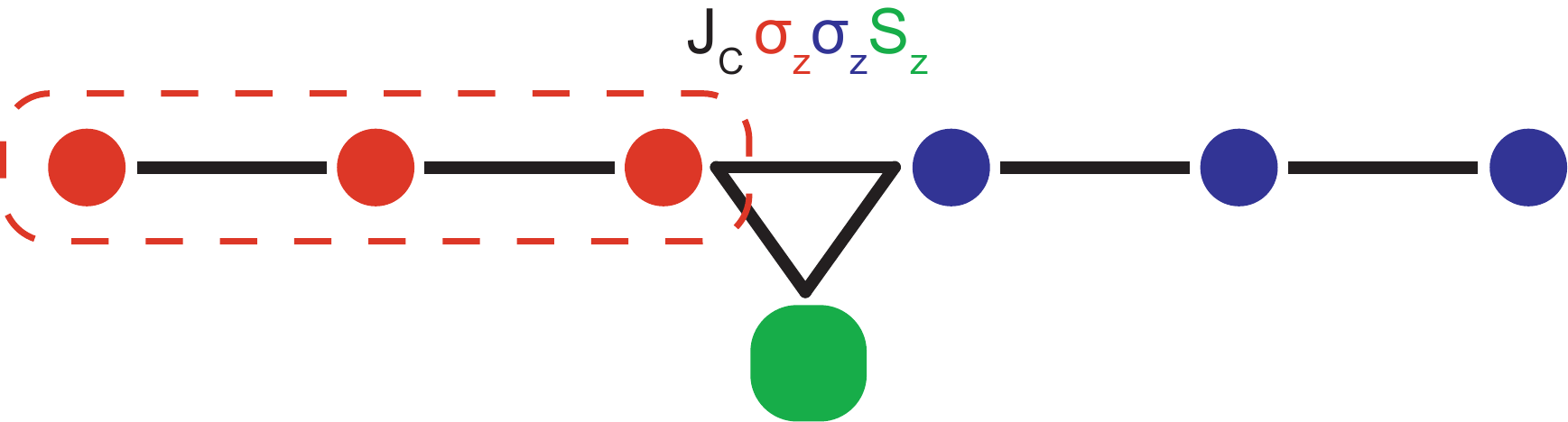}}
    \caption{Minimal system on which braiding properties can be observed. An extraneous coupler spin connects two identical Ising chains through a 3-spin interaction.}
    \label{fig_ising_braiding}
\end{figure}

For the particular case of Ising chains, Backens et al. have shown\cite{backens_emulating_2017} that with two Ising chains connected by a coupler and through a rotation of the coupler midway during the braiding procedure, it is possible to operate on a 1D chain. In the following, we will focus on this case, presented in Fig.~\ref{fig_ising_braiding}, which is described by the following Hamiltonian:

\begin{equation}
    \label{eq:chain}
    H = \sum_{\alpha} H_{Ising, \alpha} + H_{CI}
\end{equation}
    
\begin{equation}
    H_{CI}=-J_{C}S^{z}\sigma_{1}^{z}(-1)\sigma_{2}^{z}(1)
    \label{eq:coupler}
\end{equation}
    
where $\sigma_{\alpha}^{z}(-1)$ and $\sigma_{\alpha}^{z}(1)$ corresponds to the last and first site of a chain $\alpha$, respectively, with $H_{CI}$ representing the coupler interaction.

\begin{figure}[ht!]
    \centerline{\includegraphics[width=0.48\textwidth]{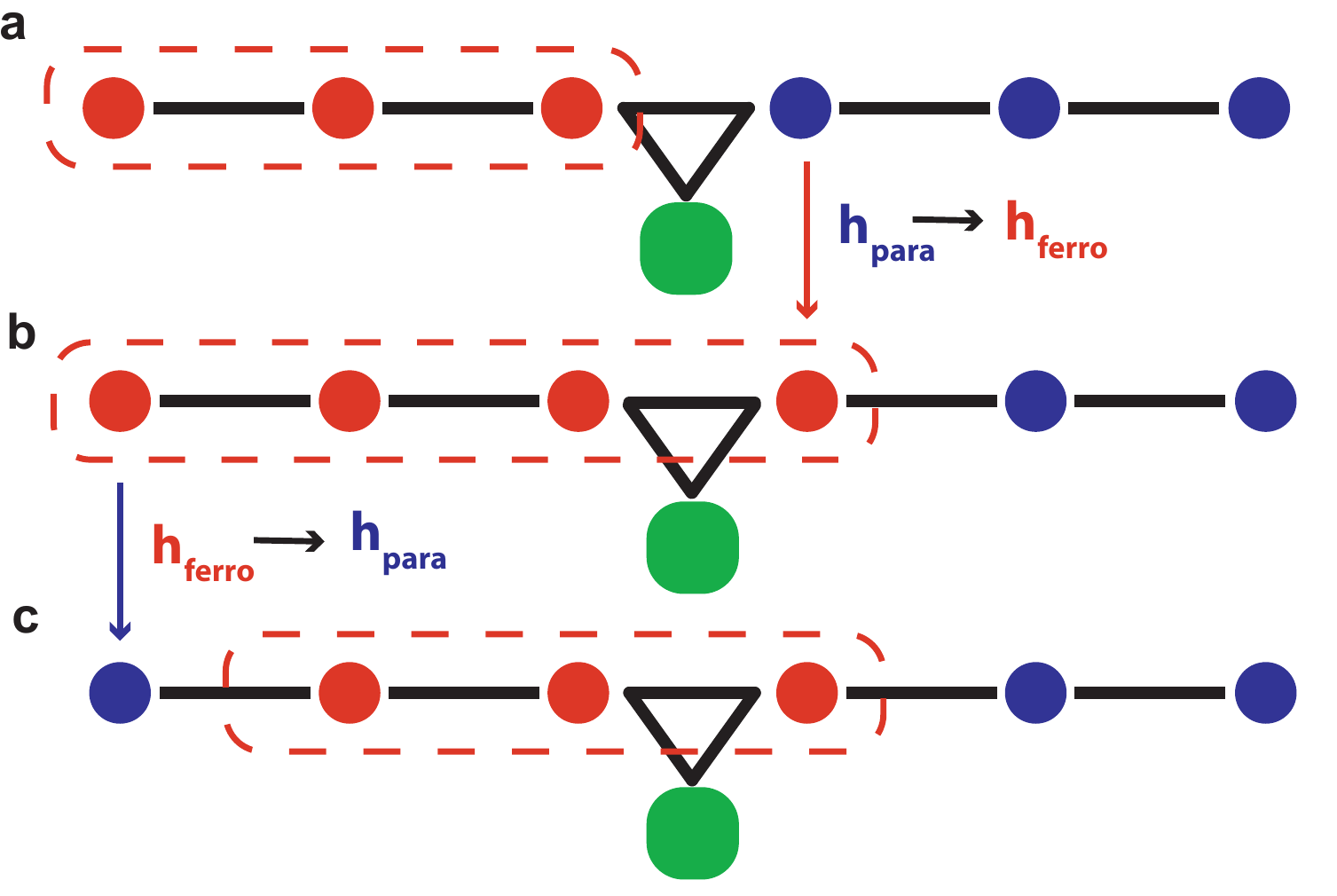}}
    \caption{Stages involved in moving the ferromagnetic domain. Starting from configuration \textbf{a}, the field on the fourth site is ramped down adiabatically from $h_{para}$ to $h_{ferro}$. As a consequence in \textbf{b}, the ferromagnetic region spans 4 sites. Ramping the field up on the leftmost site, the ferromagnetic region is contracted by one site as shown in \textbf{c}.}
    \label{fig_2}
\end{figure}

In addition to the coupler manipulation, the braiding procedure involves moving the ferromagnetic region along the chain. This can be achieved through manipulation of local fields $h_{n}$. It is important here that each step of the procedure is done slowly enough, in an adiabatic fashion, so that the system remains in its ground state. By adiabatically decreasing $h_{n}$ on a site adjacent to the ferromagnetic domain from $h_{para}$ to $h_{ferro}$, the ferromagnetic domain can be elongated by one site from right (left) \cite{dorner_entangling_2003}. The reverse process can be used to shorten the domain from left (right). Consequently, the ferromagnetic region can be moved along the chain by successively elongating the ferromagnetic domain from one end and then shortening from the other end, as shown in Fig.~\ref{fig_2}.

\begin{figure}[ht!]
    \centerline{\includegraphics[width=0.48\textwidth]{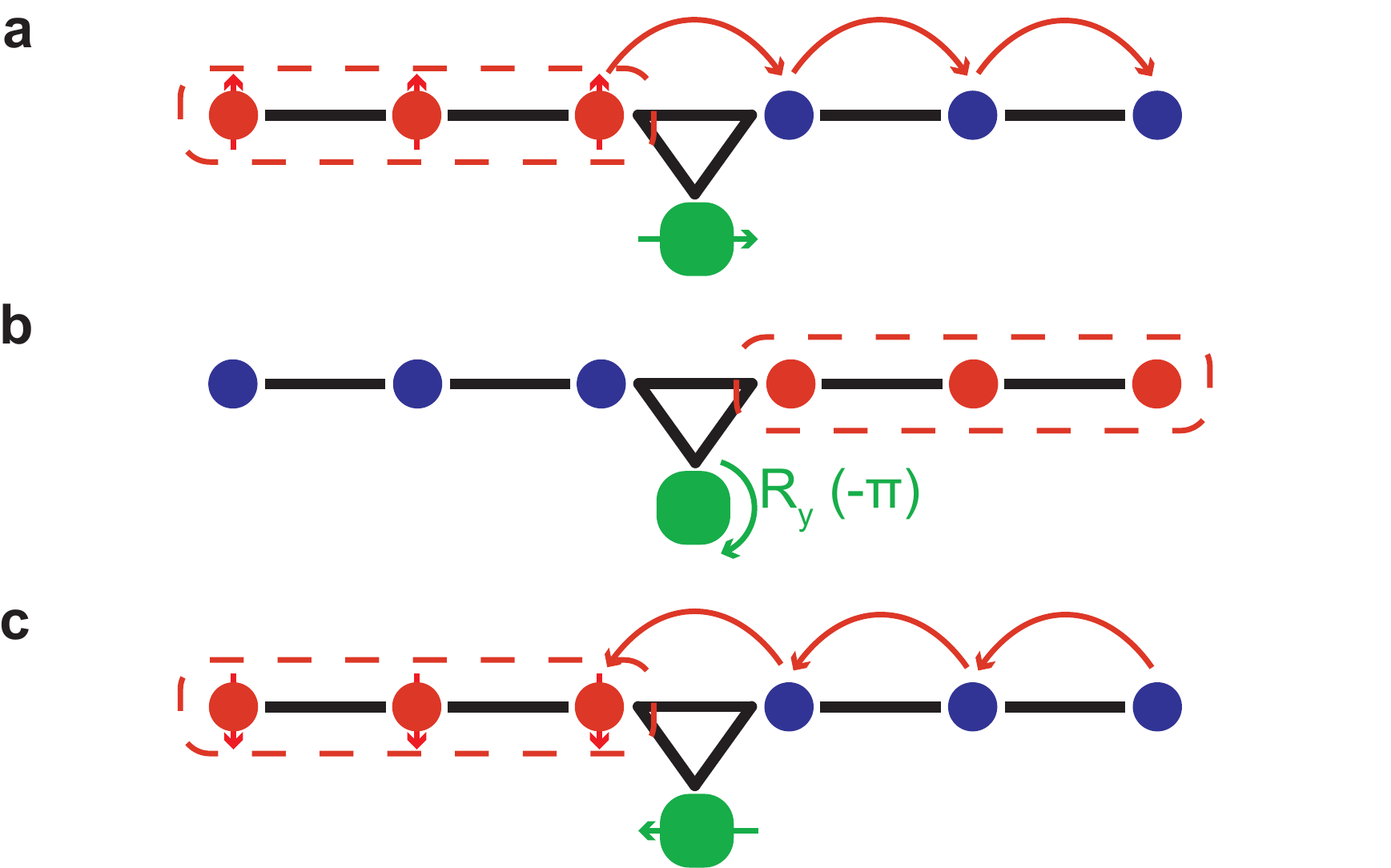}}
    \caption{Braiding procedure on an Ising chain. \textbf{a} The ferromagnetic region on the left is first initialized in $\ket{\uparrow\uparrow\uparrow} = \frac{1}{\sqrt{2}}\left(\ket{0} + \ket{1}\right)$ and the coupler is initialized in $\frac{1}{\sqrt{2}}\left(\ket{\uparrow} + \ket{\downarrow}\right)$. The ferromagnetic region is then transported adiabatically to the right side of the chain. \textbf{b} The coupler spin is rotated by $\pi$ around its y-axis. \textbf{c} The ferromagnetic region is brought back to the left and ends up in $\ket{\downarrow\downarrow\downarrow} = \frac{1}{\sqrt{2}}\left(\ket{0} - \ket{1}\right)$ as a result of the braiding which behaves like a $R_z(\pi)$ gate in the logical qubit space ($\ket{0},\ket{1}$).}
    \label{fig_3}
\end{figure}

The complete braiding process is described in Fig.~\ref{fig_3}, starting with the ferromagnetic domain on the far end of the left chain. The coupler is rotated around the x-axis by $\pi/2$ so that it is in a superposition of $S^{z}=+1$ and $S^{z}=-1$. The ferromagnetic domain is then moved adiabatically, as described above, towards the far end of the right chain. On the way, it interacts with the coupler. The coupler is then rotated around the y-axis with angle $\theta$, and the ferromagnetic domain is moved adiabatically back to the far end of the left chain. 

The overall braiding operation induces a rotation of $-\phi$ around the z-axis of the $(\ket{0}, \ket{1})$ subspace, whose states are given in Eq.~\ref{eq:states}. In particular, if we initialize the ferromagnetic region in $\ket{\uparrow\uparrow\uparrow}$ and perform of rotation of $\phi = \pi$ the final state of the system is expected to be $\ket{\downarrow\downarrow\downarrow}$.

\section{\label{sec:braid}Digital Simulation of Braiding in 1D}

The system's time evolution is described by a circuit of gate operations in the qubit space \cite{vatan_realization_2004,santos_reverse_2018}. In this case, we map each spin in the Ising chain to a qubit and use an extra qubit to describe the coupler required by the protocol since we are dealing with spin-1/2s, which are essentially qubits. Simulating the braiding protocol described in \ref{ss:ising.braiding} requires simulating the Hamiltonian dynamics of the system and performing adiabatic changes to its parameter. Next, we discuss each aspect and introduce the parameters that can be used to optimize the procedure.

The time evolution corresponding to the Hamiltonian of the system during a time interval $\Delta t$ is written as $U = \exp(-iH\Delta t)$. 
Since the circuit implementation is limited to single- and two-qubit gates, the Hamiltonian is decomposed into a minimal set of four terms, i.e. $H=\sum_X H_X$, that can be exponentiated efficiently, that is with a fixed number of gates irrespective of the system size with cost $O(1)$.
These summands are the nearest-neighbour coupling interaction for even sites $H_{J,e}=-J\sum_{n=2k}{\sigma^z_n\sigma^z_{n+1}}$ as well as odd ones $H_{J,o}=-J\sum_{n=2k+1}{\sigma^z_n \sigma^z_{n+1}}$, the Zeeman terms $H_Z=-\sum_{n}h_n\sigma^x_n$, and the coupler interaction term $H_{CI} = -J_C S^z\sigma^z_1(1)\sigma^z_2(1)$.
Assuming that the time step is sufficiently small, i.e., $h\Delta t\ll 1$, one can approximate the time evolution operator as $\prod_X \exp(-iH_X\Delta t)$ using the Suzuki-Trotter expansion.
The adiabatic condition and the Trotter error bounds are discussed in the next section. 

The nearest-neighbor interaction terms of the Ising Hamiltonian, $J\sigma^{z}_{n}\sigma^{z}_{n+1}$ described in Eq.~\ref{eq:ising}, can be realized as follow. First, a controlled-NOT (CNOT) gate is added between qubit $q_{n}$ (control) and $q_{n+1}$ (target). We then use an $R_{z}(J\,\Delta t)$ gate to rotate $q_{n}$ around the z-axis. Finally, we apply another CNOT gate between $q_{n}$ (control) and $q_{n+1}$ (target) to complete the time evolution of the nearest-neighbor interaction, i.e. $\exp(-iJ\sigma_n^z\sigma_{n+1}^z\Delta t)$.
The on-site magnetic field, $h_{n}\sigma^{x}_{n}$, the second term of Eq.~\ref{eq:ising}, is implemented using a single-qubit rotation in the $x$ direction \cite{vatan_realization_2004,santos_reverse_2018} as $\exp(-ih_n\sigma_n^x\Delta t) \equiv R_{x}(h_{n}\,\Delta t)$.

\begin{figure}[ht!]
    \input{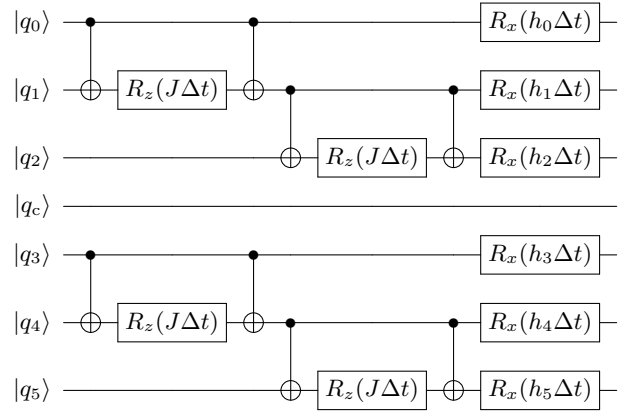}
    \caption{Quantum circuit representation of the Ising chain Hamiltonian with the nearest-neighbor interaction displayed on the left and the magnetic field interaction displayed on the right.}
    \label{fig_4}
\end{figure}

The circuit describing the time evolution of $H_{j,e}$, $H_{j,o}$, and $H_Z$, for two chains of size three, is presented in Fig.~\ref{fig_4}. Note that to reduce the depth of the circuit, we apply the pair interaction to the $q_{n},q_{n+1}$ pair and to the $q_{n+2},q_{n+3}$ pair simultaneously, allowing to describe the complete interaction of the chain with only two layers of gates.

\begin{figure}[ht!]
    {\input{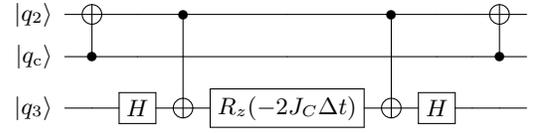}}
    \caption{Quantum circuit representation of the coupler interaction with two qubits, representing the ends of the chains, where the coupler is the middle qubit denoted by $|q_\text{c}\rangle$.}
    \label{fig_5}
\end{figure}

Finally, To implement the time evolution of the coupler interaction seen in Eq.~\ref{eq:coupler}, i.e., $\exp(-iJ_{C}S^{z}\sigma^{z}_1(1)\sigma^{z}_{2}(1)\Delta t)$, we use a combination of CNOT, Hadamard and $R_{z}$ gates as shown in Fig.~\ref{fig_5} as described in \cite{vatan_realization_2004}. 

We present the simulation of 1D braiding with a rotating coupler representing a phase gate, $R_z(\pi)$, in the basis of the logical states defined in Eq.~\ref{eq:states}. The braiding process involves four stages:

\subsubsection{Initialization of Chain and Coupler}

We begin by creating a circuit of qubits with one qubit for each site and an extra qubit for the coupler. We initialize the ferromagnetic domain in either one of the defined two logical states: $\ket{0}$, $\ket{1}$, or in one of their superpositions. To start from a logical state, we use a series of Hadamard, CNOT, and $R_{y}(-\frac{\pi}{2})$ gates applied to the qubits of the ferromagnetic domain. Regarding the superpositions, the $\ket{\uparrow\uparrow\uparrow}$ is trivially accessible as the system's default state. Hadamard gates are applied to all the qubits in the paramagnetic domain, and the coupler is initialized by applying an $R_{x}(\frac{\pi}{2})$ gate on the qubit.

The initialization sequence relies on the fact that the chosen value of $h$ in the ferromagnetic domain, $h_{ferro}$, is much smaller than $J$, making the initial state a good approximation of the ground state. A similar argument applies to $h_{para}$ used in the paramagnetic domain.

\subsubsection{Moving Ferromagnetic Domain from Left to Right}
\label{dig.braid.ltr}

Given the initial configuration where the ferromagnetic domain is on the far left and the final configuration is on the far right, the intermediate field configurations are determined. To move the domain one way or the other, the domain is extended by one site on the side closer to the direction of the movement and contracted by one site on the other side, as seen in Fig. \ref{fig_2}. In contrast to what was discussed in \ref{ss:ising.braiding}, we perform the extension and the contraction of the ferromagnetic domain simultaneously rather than sequentially. We have verified that this does not degrade the fidelity of the process and allows for shallower circuits. As a consequence, each intermediate configuration represents the ferromagnetic domain's movement by one site.
The on-site fields are updated in small steps between each configuration, where the Trotter expansion is used to describe the evolution. We chose a simple linear update for the on-site field, which in some cases could suffer from the large values required for $h_{para}$. The size and frequency of the field update affect the adiabaticity of the evolution.

\subsubsection{Mid-Braiding Manipulation}

After the ferromagnetic domain is moved to one side, the coupler is rotated around the y-axis using an $R_{y}$ gate in a series of steps. With each rotation step, the site's interaction with the slightly rotated coupler is represented using the Trotter evolution. This approach assumes that the coupler rotation is not much faster than the Hamiltonian evolution.

\subsubsection{Moving Ferromagnetic Domain Back from Right to Left}

Following the moving protocol described in \ref{dig.braid.ltr}, the ferromagnetic domain is moved from right to left. This concludes the $R_z(\pi)$ braiding process for which an initial state of $\ket{\uparrow\uparrow\uparrow}$ and a total rotation of the coupler by $\pi$ should lead to $\ket{\downarrow\downarrow\downarrow}$.

\label{sim-par}
Based on the description of the simulation provided above, we can identify the following relevant parameters that could be susceptible to influence the fidelity of the procedure:

\begin{itemize}
    \item Trotter step - $\Delta t$: the duration for which we evolve the system in each iteration of the Trotter procedure. Using shorter steps should increase the approximation accuracy but would increase the overall depth of the circuit.
    \item Phase separation: represents how well the ground states of the ferromagnetic and paramagnetic region, defined by the simple states used in the initialization step, depend on $h_{ferro}/J$ and $h_{para}/J$ which need to be, respectively, much smaller and much larger than unity. Since the procedure needs to remain in the system's degenerate ground state, it is important that this approximation is accurate.
    \item Field update rate - $\Delta h/T$: the update of the on-site magnetic field when converting a site needs to be adiabatic. Consequently, we expect slower updates to give better fidelity. Since we perform step updates, this parameter is controlled by both the magnitude of the update ($\Delta h$) and their temporal distance ($T$). There may exist an optimal choice of those two for a given rate. In particular, large steps are unlikely to yield accurate results.
    \item Coupler interaction strength - $J_C$: The coupling strength between the coupler and the two chains, which corresponds to the chemical potential in a Kitaev chain, may influence the braiding's fidelity. In particular, the back-action of the coupler rotation on the adjacent ferromagnetic domain may be mitigated by lower coupling strength.
    \item Coupler rotation speed - $\Gamma$: how fast the coupler's rotation is performed during the mid-braiding manipulation, after moving the ferromagnetic domain from left to right, may influence the fidelity of the braiding process.
\end{itemize}
\section{Simulation Performance and Fidelity}

 Here we present the results of implementing the phase gate, $R_z(\pi)$ in the $\{\ket{0}, \ket{1}\}$ subspace, using the braiding protocol discussed in Section \ref{sec:braid}. To investigate the performance of the Hamiltonian simulation and the simulation's fidelity, $F$, we start by looking at five different situations and measure the fidelity in each case on a chain of 6 sites, $N_{s}=6$,  with a ferromagnetic domain of 3 sites:
 
 \begin{itemize}
     \item Two separate Ising chains with no coupler spin and simply performing the movement of the ferromagnetic region from the left to the right and back. For this operation, we start with the ferromagnetic domain initialized either in $\ket{\uparrow\uparrow\uparrow}$ or $\ket{0}$.
     \item Two Ising chains connected through a coupler and performing the same operation as above with the coupler initialized along the +$x$-axis, as in the braiding protocol. We consider again two initial states: $\ket{\uparrow\uparrow\uparrow}$ or $\ket{0}$.
     \item The full braiding applied on $\ket{\uparrow\uparrow\uparrow}$ which involves the coupler's rotation after the first translation.
 \end{itemize}
 
 In each case above, except for the braiding, we expect to recover the original state at the end of the procedure. All the following quantum circuit simulations were run on IBM's Qiskit Aer QASM Simulator with 10000 shots. The QASM simulator acts as an ideal quantum computer exhibiting only sampling errors. Another main source of error in the simulation is attributed to the termination of the Trotter expansion as will be discussed. Other types of errors that arise from NISQ hardware are discussed also in the last section. 
 Through a process of optimization starting with reasonable estimates, the optimized parameters, reported in Table \ref{tab:op_par}, are obtained and used for all simulations where we consider units such that $\hbar$ = 1. The nearest-neighbour interaction strength is always set at $J=1$ throughout all the simulations. 

\begin{table}[htbp]
    \caption{Parameter Values for Shallowest Circuit Depth for $N_{s}=6$} 
    \begin{center}
        \begin{tabular}{|c|c|c|c|c|c|c|c|}
        \hline
        Fidelity & $\Delta t$ & $h_{ferro}$ & $h_{para}$ & $\Delta h$ & T & $J_C$ & $\Gamma$\\
        \hline
        $>99\%$ & 0.2 & 0.01 & 5.0 & 0.05 & 2& 0.3 & $\pi$/3\\
        \hline
        $>90\%$  & 0.7 & 0.01 & 1.5 & 0.1 & 2 & 0.3 & $\pi$/2\\
        \hline
        \end{tabular}
    \label{tab:op_par}
    \end{center}
\end{table}

\subsection{Parameter Dependence and Optimization}

\begin{figure}[ht!]
    \centering
    \includegraphics[width=0.48\textwidth]{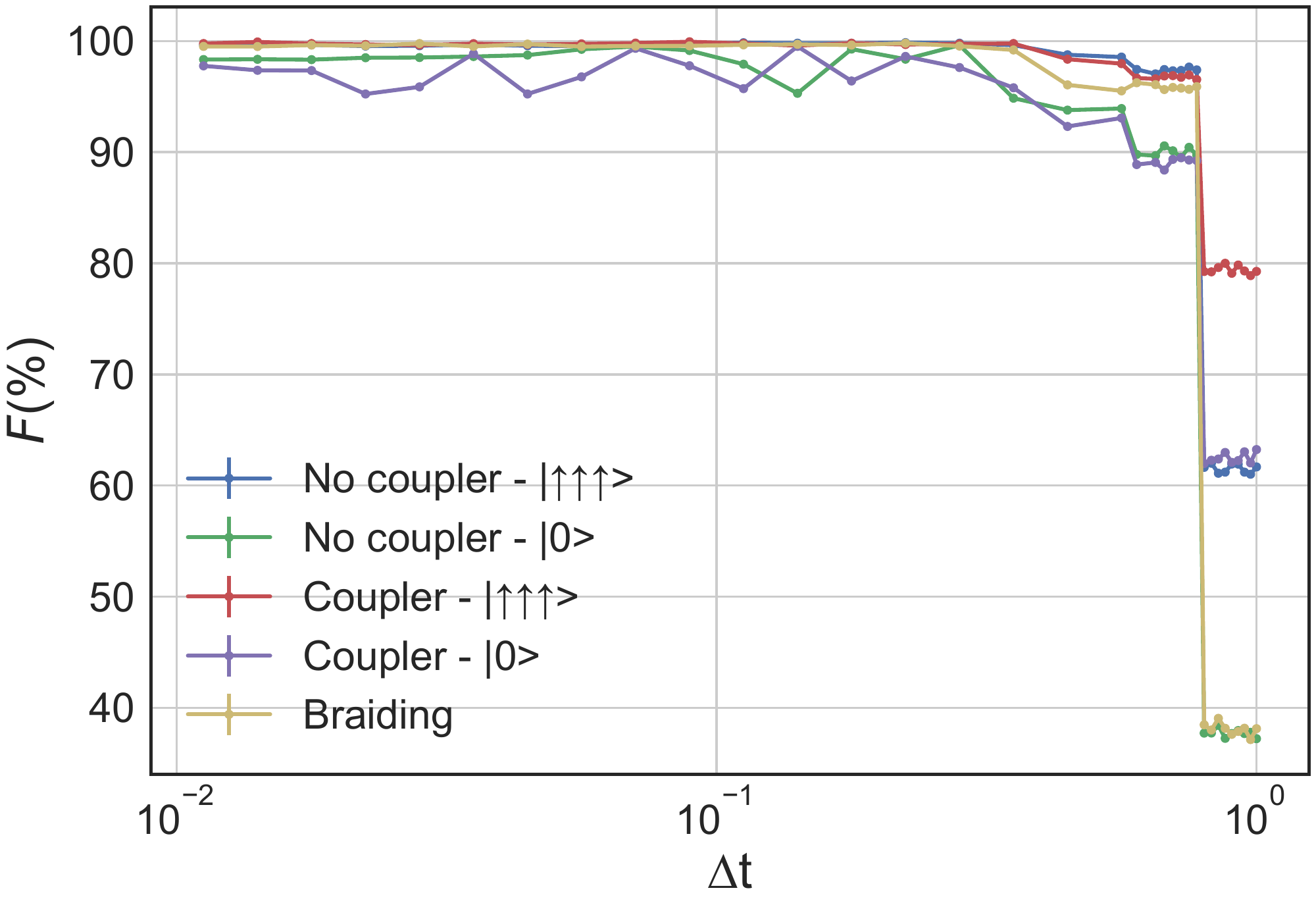}
    \caption{Fidelity, $F$, as a function of the Trotter time step, $\Delta t$ in the presence and absence of the coupler interaction for different initial states as well as the full braiding procedure. The parameters that not varied are taken from Table \ref{tab:op_par}.}
    \label{fig:del_t}
\end{figure}

\begin{figure}[ht!]
    \centering
    \includegraphics[width=0.48\textwidth]{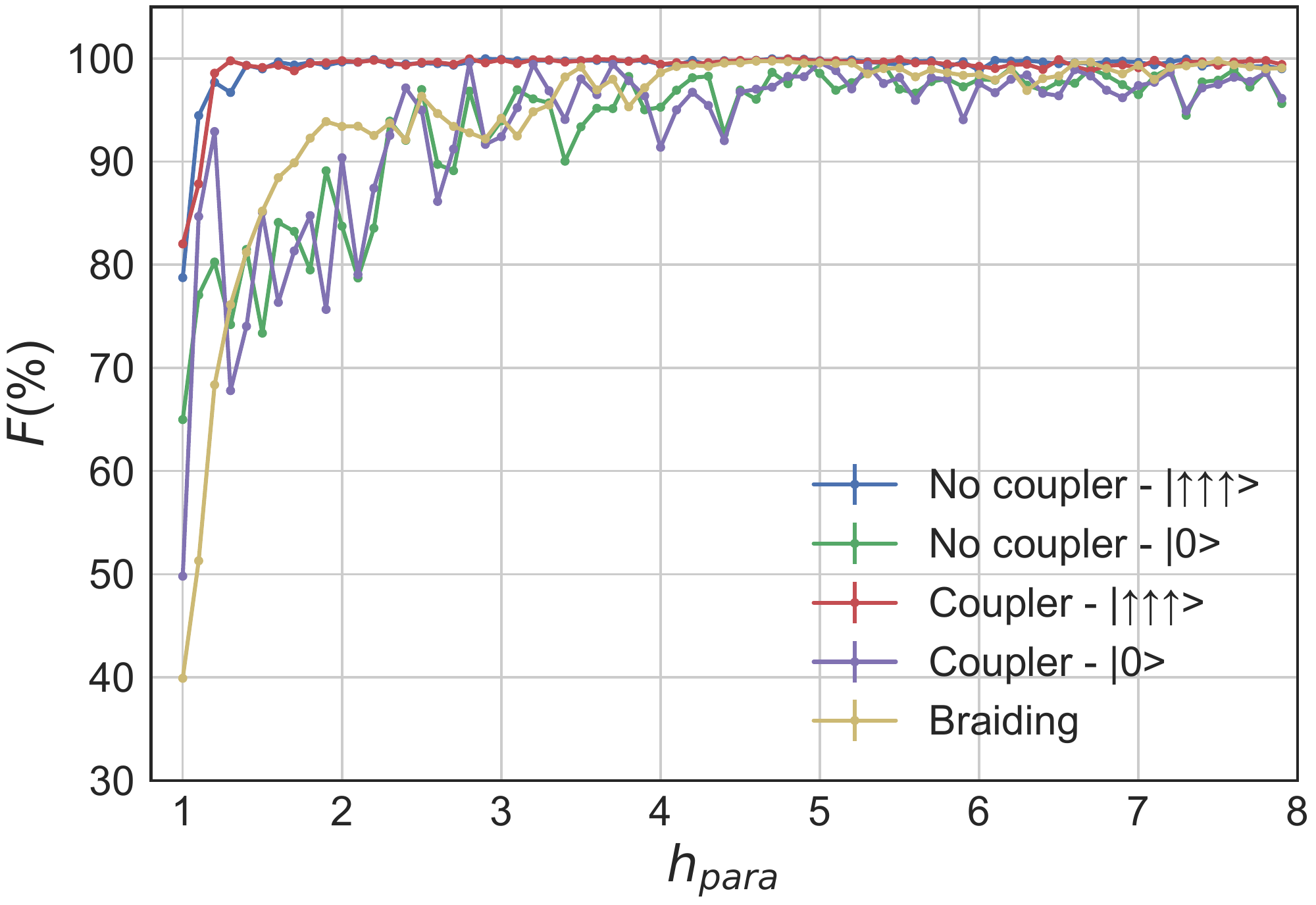}
    \caption{Fidelity, $F$, as a function of the paramagnetic field ,$h_{para}$, in the presence and absence of the coupler interaction for different initial states as well as the full braiding procedure. The parameters that are not varied are taken from Table \ref{tab:op_par}.}
    \label{fig:h_para}
\end{figure}

\begin{figure}[ht!]
    \centering
    \includegraphics[width=0.48\textwidth]{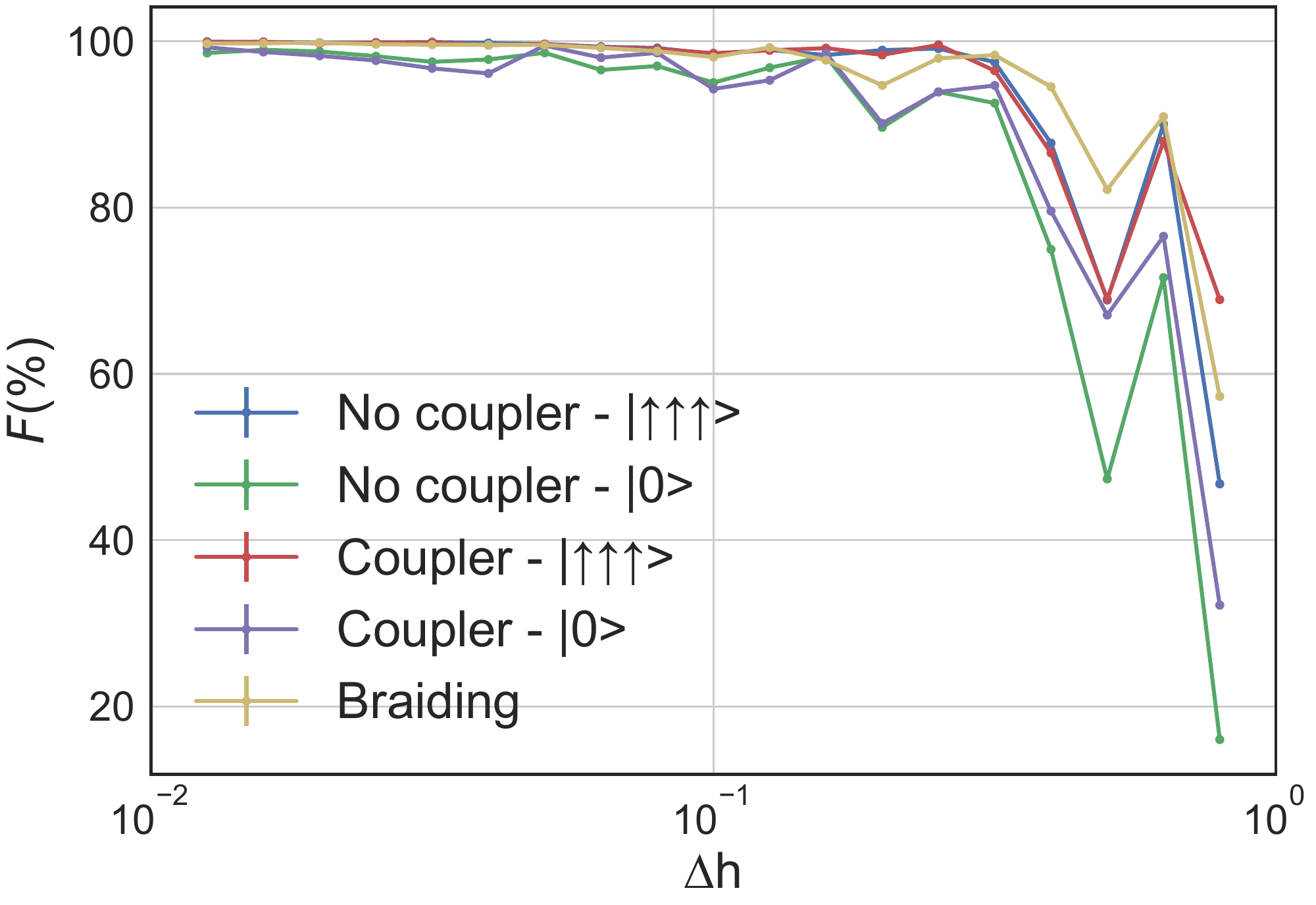}
    \caption{Fidelity as a function of the field update rate, $\Delta h$, in the presence and absence of the coupler interaction for different initial states as well as the full braiding procedure. The parameters that are not varied are taken from Table \ref{tab:op_par}.}
    \label{fig:del_h}
\end{figure}

\begin{figure}[ht!]
    \centering
    \includegraphics[width=0.48\textwidth]{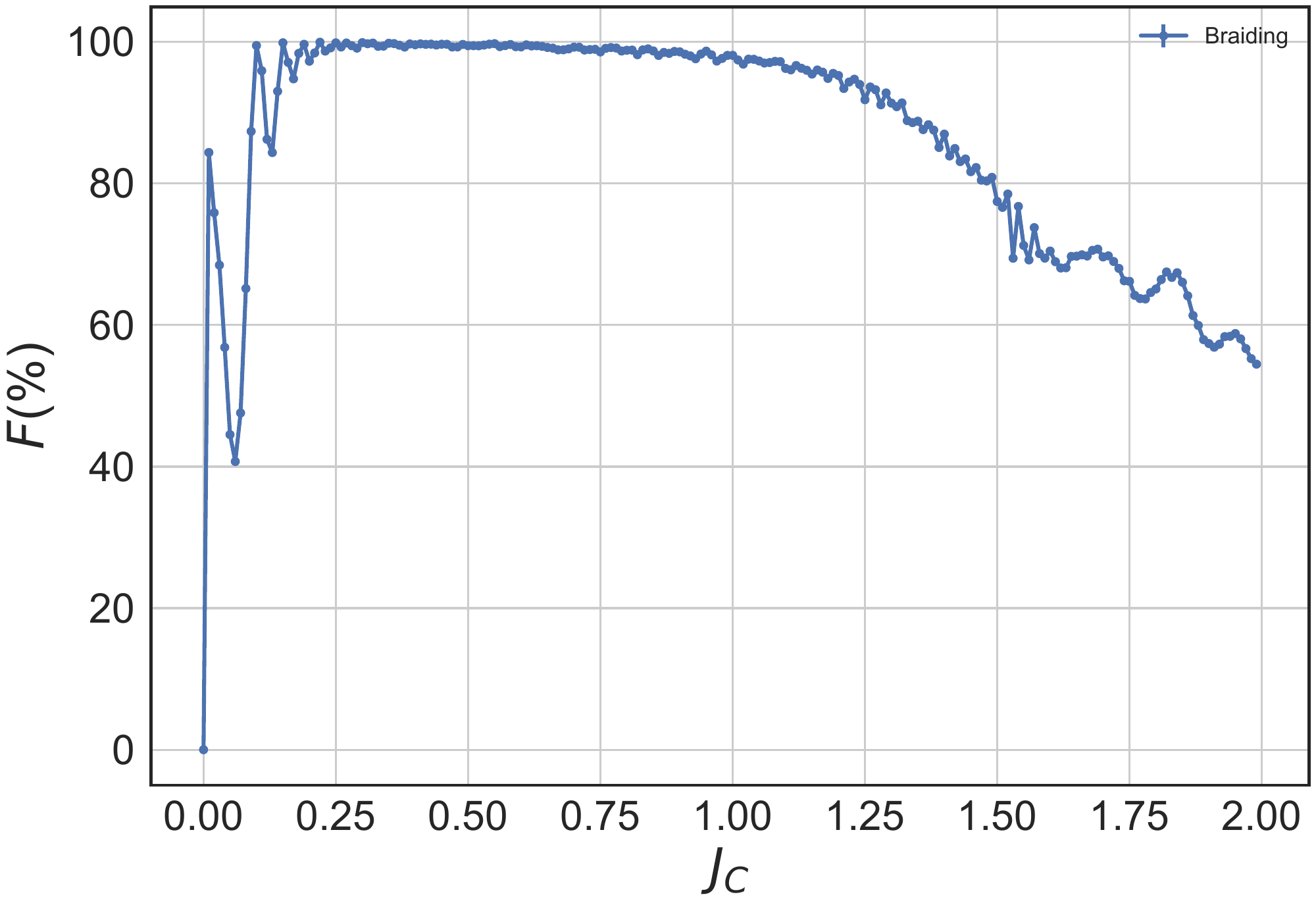}
    \caption{Braiding Fidelity as a function of the coupler interaction strength $J_C$. The parameters that are not varied are taken from Table \ref{tab:op_par}.}
    \label{fig:CI}
\end{figure}

Results for the impact of different system and evolution parameters, i.e. trotter time step size($\Delta t$), value of the magnetic field on sites in the paramagnetic domain($h_{para}$), field update rate($\Delta h$) and the strength of the coupler interaction ($J_{C}$), on the simulation's fidelity are presented in Figs.~\ref{fig:del_t}-\ref{fig:CI}. Different curves correspond to each of the five situations described above to show the impact of the initial state and the presence of the coupler on the fidelity of the operations.

Fig.~\ref{fig:del_t} presents the Fidelity $F$ scales inversely with the size of the Trotter time step, $\Delta t$ as expected.
The drop in the fidelity can generally arise from two different approximations; the adiabatic condition and the truncation error of the Trotter expansion. The adiabatic condition states $\Delta t \ll (2J)T/\Delta h$. As seen from Table \ref{tab:op_par}, the optimized parameters are sufficiently far from being non-adiabatic and, therefore, the drop in the fidelity is mostly related to the truncation error of the Trotter expansion. As seen from Eq. \ref{eq:error-bound} the error is proportional to $\Delta t$. The sudden drop in the fidelity can therefore be attributed to the non-commutativity of the summands of the Hamiltonian which becomes more pronounced as the time step increases. Similarly, the dependence of the fidelity on the initial state can be understood through the Trotter expansion. The leading Trotter's error is proportional to the commutators of the summands of the Hamiltonian which, in turn, have different expectation values depending on the initial states.

The dependence of the fidelity on the Zeeman field of the paramagnetic domain, $h_{para}$, is depicted in Fig. \ref{fig:h_para}. At small values of the field, i.e. $h_{para} < 2$, the fidelity drops drastically, especially for braiding. This is due to the fact that a successful translation and braiding operation requires a clear distinction between the ferromagnetic and paramagnetic domains which is achieved through a high Zeeman field, i.e. $h_{para} \gg J$. This can also be understood through the fact that the Zeeman field in the Ising model is equivalent to the chemical potential in the Kitaev model \cite{backens_emulating_2017} where it is used to define trivial/non-trivial regions locally. Interestingly, even though the energy difference between the states $\ket{\uparrow\uparrow\uparrow\leftarrow\leftarrow\leftarrow}$ and $\ket{\uparrow\uparrow\leftarrow\leftarrow\leftarrow\leftarrow}$ is $|J-h|$, which if the middle spin is in a ferromagnetic region is about $J$, stabilizing the paramagnetic region requires fields leading to a much larger energy separation. Conversely, for a very large value of the field, $h_{para}>50$, the fidelity starts to drop (not shown in the figure). The reason is that the Trotter error is quadratic in $h_{para}$ since $h_{para}$ appears in the commutators, $[H_{J,e}, H_Z]$, $[H_{J,o}, H_Z]$, $[H_{Z}, H_{CI}]$, and also in the total number of Trotter steps.

We further examine the fidelity's dependence on the field update increments $\Delta h$, which is proportional to the update rate $\Delta h/T$, in Fig.~\ref{fig:del_h}. Optimum fidelities are achieved only for update rates $\Delta h/T$ smaller than 0.25. The drop in the fidelity at $\Delta h > 0.25$ suggests the onset of the violation of the adiabatic condition as the update rate becomes comparable to the excitation gap of the system. The Trotter error is inversely proportional to $\Delta h$ because smaller update rate corresponds to a larger number of steps. However, this asymptotic behaviour is not observed for the values shown in the Fig.~\ref{fig:del_h}. 

Lastly, we investigate the effect of the coupler interaction strength $J_C$ on the fidelity. Fig.~\ref{fig:CI} shows the fidelity dropping at lower values, i.e., $J_C < 0.75$, which can be understood as the breaking of the ferromagnetic/anti-ferromagnetic interaction mediated by the coupler which drives the braiding process.
As the coupling between the chains is weakened, it gets more difficult for the ground state to rotate in the degenerate subspace and therefore degrades the braiding process. 
For large values of $J_C$ where it becomes considerable compared to $J$,  the fidelity drops as well. As discussed in the previous section, this may be related to the back-action of the coupler rotation. This is also consistent with the Trotter error containing the commutator $[H_Z, H_{CI}]$ which is proportional to $J_C$.
The optimal value of the coupling strength is therefore chosen according to Fig.~\ref{fig:CI} from the interval $0.25 < J_C < 0.75$. 

The coupler rotation speed $\Gamma$ appeared to only minimally affect the fidelity of the braiding operation in our simulations. This is interesting since Ref. \cite{backens_emulating_2017} suggests only to manipulate the coupler when the ferromagnetic domain is far from the coupler, and here we find that even when the domain is adjacent to the coupler, in a chain of 6 sites and a ferromagnetic domain of 3 sites, relatively fast manipulations remain possible without losing fidelity.

\subsection{Circuit Depth}

\begin{figure}[ht!]
    \centerline{\includegraphics[width=0.5\textwidth]{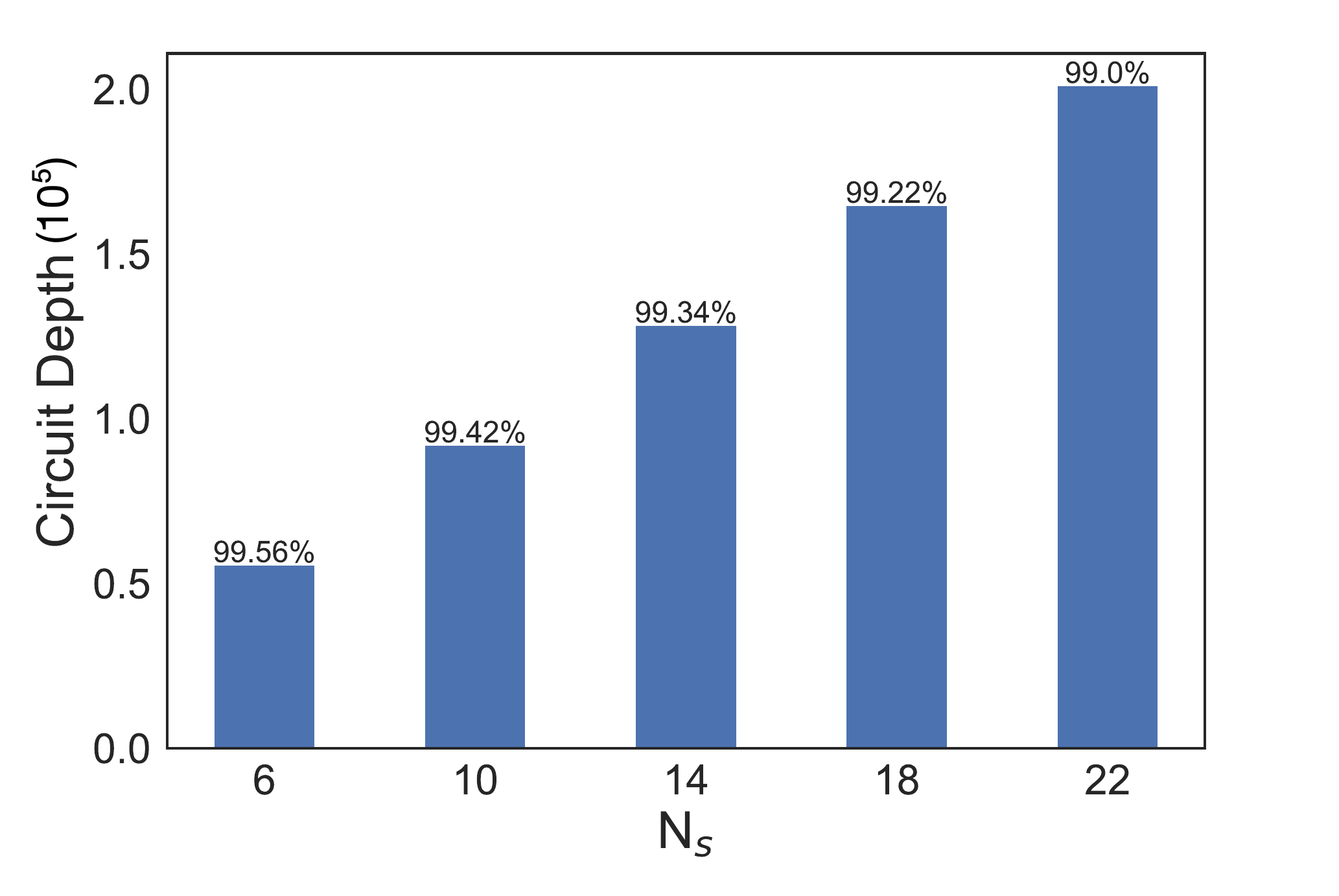}}
    \caption{The circuit depth, defined as the number of non-simultaneous gates executed, for optimum braiding fidelity, $F$, as a function of the size of the system. The optimum fidelity achieved is noted above the bars, and the ferromagnetic domain's size is half the system's size.}
    \label{depth_size}
\end{figure}

The depth of the circuit can be estimated in terms of the simulation parameters. 
According to the circuits in Figs. \ref{fig_4} and \ref{fig_5}, the total circuit depth of each Trotter time step that includes the full Hamiltonian interaction is $D_T = 12$, as seen from the number of non-simultaneous gates and calculated by the Qiskit simulator. Shifting the ferromagnetic domain by one site requires $h_{para}/\Delta h$ field updates each with $T/\Delta t$ Trotter steps.
The rotation of the coupler requires  $\pi/\Gamma$ updates for each Trotter step. 
Therefore, the braiding operation, i.e., shifting the ferromagnetic domain by $N_s$ sites along with the coupler rotation, results in the circuit depth 
\begin{equation}
    D = D_T(\frac{T}{\Delta t})\bigg(N_s(\frac{h_{para}}{\Delta h}) + \frac{\pi}{\Gamma}\bigg). 
\end{equation}
Here $N_s$ is the total number of Ising spins in the system. 
We note that this is an upper bound to the depth, and a shallower circuit can be obtained in practice by combining commuting gates.

For a given fidelity $F$ and system size $N_{s}$, we minimize the circuit depth by examining $T/\Delta t$, $h_{para}/\Delta h$, and $\pi/\Gamma$. A braiding fidelity of $F>99\%$ can be achieved through optimization with the depth being as low as 56518 with parameter values shown in Table \ref{tab:op_par}. Using our results, we also found parameters, detailed in Table \ref{tab:op_par}, yielding above $90\%$ fidelity with a depth reduced to 2679 which is significantly more accessible to the current NISQ hardware.

For larger systems, we investigated the depth while maintaining  $F\approx99\%$ for a system of size up to 22 sites, corresponding to a ferromagnetic domain of 11 sites. The circuit depth is seen to increase linearly with the size of the system reaching a circuit depth above 200k for a system of 22 sites while the fidelity slightly drops as expected, as seen in Fig.~\ref{depth_size}.

\section{Error Analysis}

\subsection{Suzuki-Trotter Error}
The time evolution of the system during the time period $T$ can be decomposed as $U(0, T)=U(T-\Delta t, T)\cdots U(\Delta t, 2 \Delta t)U(0, \Delta t)$, where each shorter time evolution is written as a time-ordered exponential 
\begin{equation}
    U(t_j, t_j + \Delta t) = \mathcal{T} \exp\bigg(-i\int_{t_j}^{t_j + \Delta t} dt H(t)\bigg).
\end{equation}
Here $H(t)$ is the time-dependent Hamiltonian of the system. 
During a time step $\Delta t$, the field at the ends of the chain changes by $\Delta h/(T/\Delta t)$ which should be sufficiently smaller than the excitation gap, i.e., $2J$, to make sure one is in the \textit{adiabatic} regime, that is $\Delta t \ll (2J)T/\Delta h$. 
When this condition holds, the integral above can be replaced with $H(t_j) \Delta t$.
Using the Baker-Campbell–Hausdorff formula along with the decomposition $H(t_j) = \sum_X H_X$, one obtains 
\begin{equation}
\begin{split}
    & U'(\Delta t) = \\
    & \bigg( \prod_X e^{-i H_X \Delta t} \bigg) 
    \bigg (e^{-\frac{1}{2}\Delta t^2\sum_{X, X'} [H_X, H_X'] + O(\Delta t^3)}\bigg), 
\end{split}
\end{equation}
where $U''(\Delta t) = \prod_X \exp(-i H_X \Delta t)$ is the product formula implemented in the quantum circuit. 
The leading error is written as  \cite{poulin2011quantum,huyghebaert1990product}
\begin{equation}
\label{eq:error-exp}
    \lVert U'(\Delta t) - U''(\Delta t) \rVert \le \frac{1}{2}\sum_{X, X'}\lVert[H_X, H_{X'}]\rVert (\Delta t)^2. 
\end{equation} 
For the Hamiltonian considered here only three of these commutators are non-zero, i.e., 
\begin{equation}
\begin{split} 
& [H_{J,e}, H_Z]=\sum_{j=2k}2iJ(h_j\sigma_j^y \sigma_{j+1}^z + h_{j+1}\sigma_j^z \sigma_{j+1}^y),\\
& [H_{J,o}, H_Z]=\sum_{j=2k+1}2iJ(h_j\sigma_j^y \sigma_{j+1}^z + h_{j+1}\sigma_j^z \sigma_{j+1}^y),\\
& [H_Z, H_{CI}]=-2iJ_CS^z(h_1\sigma_1^y\sigma_2^z + h_2\sigma_1^z\sigma_2^y). 
\end{split}
\end{equation}
Since at any given time half of the sites are ferromagnetic, i.e. $h_j \approx 0$, one can find a lower upper bound for the commutators above, that is $\lVert[H_{J,e}, H_Z]\rVert + \lVert[H_{J,o}, H_Z]\rVert \le 2N_sJh_{para}$. 
The upper bound for the last commutator is obtained as $\lVert[H_Z, H_{CI}]\rVert \le 4J_Ch_{para}$.
Therefore, the leading error corresponding to each time step of the Suzuki-Trotter expansion is
\begin{equation}
\label{eq:error-bound}
    \lVert U'(\Delta t) - U''(\Delta t) \rVert \le (N_sJ + 2J_C)h_{para}(\Delta t)^2.
\end{equation}
This error corresponds to a single Trotter step during which the field is changed by $\Delta h/(T/\Delta t)$. The total error of changing the field by $\Delta h$ during time period $T$ adds a factor of $T/\Delta t$. Ramping up the field to $h_{para}$ on each site requires another factor of $h_{para}/\Delta h$. Lastly, the movement of the ferromagnetic domain from one end of the chain to the other end and back adds a factor of $N_s$ to the error bound.
Therefore, the total leading error due to the truncation of the Suzuki-Trotter expansion is 
\begin{equation}
    \lVert U' - U'' \rVert \le N_s\frac{h_{para}}{\Delta h}\frac{T}{\Delta t}(N_sJ + 2J_C)h_{para}(\Delta t)^2.
\end{equation}
This error bound demonstrates the asymptotic behaviour of the error in terms of the parameters of the simulation. 

Overall, the fidelity of the simulations seems to generally depend on both the adiabatic condition and the truncation error of the Trotter's expansion. However, plugging the default parameters into Eq. \ref{eq:error-bound} results in a loose error bound which is several orders of magnitude greater than what is seen from Figs. \ref{fig:del_t} to \ref{fig:CI}. As discussed in Refs. \cite{childs2019nearly, childs2021theory}, in a Hamiltonian with local interactions, the error may not be dominated by the second order term and therefore the true experimental error observed in the simulations are by orders of magnitude smaller than the second order error bounds. 

\subsection{NISQ Hardware Error}
In addition to errors arising from the adiabatic conditions and the Trotter expansion, which is intrinsic to the algorithm and limits the ultimate fidelity of the simulation, there are extrinsic errors that arise from the nonidealities of the hardware implementation. 
\begin{figure}[ht!]
    \centerline{\includegraphics[width=0.5\textwidth]{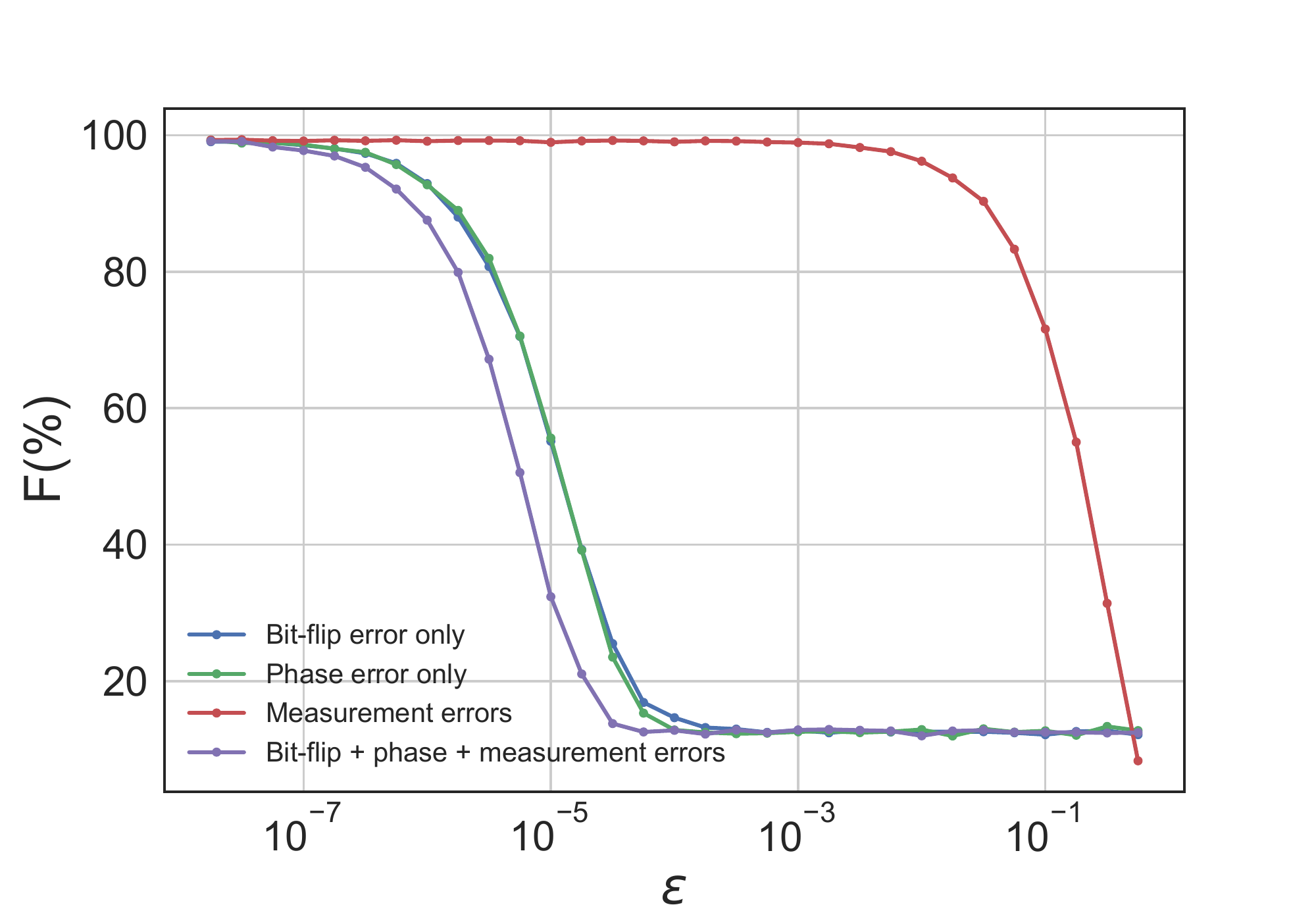}}
    \caption{Analysis of fidelity as a function of bit-flip, phase and measurement errors.}
    \label{fig:error}
\end{figure}
Using Qiskit's predefined noise models to introduce bit-flip, phase, and measurement errors on all single-qubit and two-qubit gates for all qubits we are able to estimate the simulation's tolerance to the NISQ hardware's noise. We study the optimized case where $N_{s}$ = 6 where the errors are applied to all the qubits and gates in the system. Fig. \ref{fig:error} illustrates the fidelity as a function of the error $\epsilon$ for different types of errors. 
The value of the error being applied is the same for both single-qubit and two-qubit gates. As expected, the errors are found to come mostly from two-qubit gates rather than single-qubit gates. We further observe that measurement errors alone do not have a significant effect on the fidelity as seen in Fig.~\ref{fig:error} given that the current achievable measurement errors are of the order of $\sim 10^{-2}$ \cite{tannu_mitigating_2019}. 
Phase and bit-flip errors, on the other hand, are seen to be the dominant sources of error in the system.
Achieving fidelities above 90\% requires quantum gates with an average phase and bit-flip errors of $\epsilon < 10^{-6}$. This agrees with the fact that the circuit depth is of the order of $10^{5}$ and suggests that high fidelities cannot be achieved on current NISQ hardware as the errors are several orders of magnitude greater than what is needed for a circuit depth corresponding to a single braid protocol without taking into consideration other sources of errors. 

\section{Conclusion}
In a one-dimensional system of two Ising chains, we studied a braiding-like operation that moves the wavefunction inside the degenerate subspace of the ferromagnetic state representing a logical qubit. 
The braiding operation occurs in the parameter space of the Ising chain which resembles the real space braiding of zero modes in a Kitaev chain. 
The adiabatic time evolution of the system was implemented by using the Suzuki-Trotter expansion and we achieved a fidelity of $>99\%$ for systems of up to 11 sites.
 We studied the trends of the fidelity of the simulations as a function of different system evolution parameters and obtained optimum sets of parameters for high fidelity as well as high efficiency implementations.  

The circuit depth scales linearly as a function of system size $N_{s}$, but one should note that the depth of those circuits is large. Our optimum parameters yield a depth of $\sim57000$ for $N_{s}=6$, far exceeding the capabilities of the current state of the art devices, which can handle the depth of a few 10s \cite{arute_quantum_2019}. In an effort to bridge this gap, the complexity of the system could be significantly reduced and with the parameter values detailed in Table \ref{tab:op_par}, to achieve $\sim 90\%$ fidelity with a depth reduced to about $\sim2700$. This is a considerable improvement but remains beyond present-day capabilities. 

We showed that the Trotter error bounds are, by orders of magnitude, looser than the error observed in the simulations. 
This suggests that the error might not be dominated by the second order terms in this expansion and therefore a higher fidelity with a lower circuit depth is achievable. Further, for our simulations bit-flip and phase errors contribute the most error to the simulation and a NISQ hardware needs to have an error of the order $\epsilon\sim 10^{-6}$ per gate for a optimum fidelity or above which is orders of magnitude lower than errors of quantum hardware we have access to currently.

% \appendix

% \section{Appendixes}

\bibliography{apssamp}% Produces the bibliography via BibTeX.

\end{document}